\documentclass[12pt]{article}
\pdfoutput=1

\usepackage[bulletsep]{collref}
\usepackage{amssymb,graphicx}
\usepackage[intlimits]{amsmath}
\usepackage{bbm}
\usepackage[small]{subfigure}

\usepackage{MnSymbol}

\makeatletter
\let\old@startsection=\@startsection
\let\oldl@section=\l@section
\renewcommand{\@startsection}[6]{\old@startsection{#1}{#2}{#3}{#4}{#5}{#6\mathversion{bold}}}
\renewcommand{\l@section}[2]{\oldl@section{\mathversion{bold}#1}{#2}}
\makeatother

\makeatletter
\let\old@makecaption=\@makecaption
\def\@makecaption{\small\old@makecaption}
\makeatother

\newcommand{\dc}[1]{/\!\!\!\!\, #1}

\begin{document}


\thispagestyle{empty}
\begin{flushright}\footnotesize
\texttt{NORDITA 2019-028} \\
\texttt{UUITP-9/19}
\vspace{0.6cm}
\end{flushright}

\renewcommand{\thefootnote}{\fnsymbol{footnote}}
\setcounter{footnote}{0}

\begin{center}
{\Large\textbf{\mathversion{bold} Chiral Estimate of QCD Pseudocritical Line}
\par}

\vspace{0.8cm}

\textrm{
K.~Zarembo\footnote{Also at ITEP, Moscow, Russia}}
\vspace{4mm}

\textit{Nordita, KTH Royal Institute of Technology and Stockholm University,
Roslagstullsbacken 23, SE-106 91 Stockholm, Sweden}\\
\textit{Department of Physics and Astronomy, Uppsala University\\
SE-751 08 Uppsala, Sweden}\\
\vspace{0.2cm}
\texttt{zarembo@nordita.org}

\vspace{3mm}


\par\vspace{1cm}

\textbf{Abstract} \vspace{3mm}

\begin{minipage}{13cm}
Relatively low crossover temperature suggests that chiral symmetry restoration in QCD may well be described within the low-energy effective theory. The shape of the pseudocritical line in the $T-\mu $ plane is estimated within this assumption. No critical endpoint is found for physical values of quark masses.  
\end{minipage}

\end{center}

\vspace{0.5cm}


\setcounter{page}{1}
\renewcommand{\thefootnote}{\arabic{footnote}}
\setcounter{footnote}{0}


Approximate chiral symmetry shapes, to a large extent, the low-energy properties of hadrons and interactions among them. Broken in the vacuum, chiral symmetry gets restored at sufficiently high temperature on a sharp crossover.   A real phase transition would occur if quarks were massless, and chiral symmetry were exact, or if quarks were very very light. It is believed that the crossover gets more and more pronounced at larger baryon densities and eventually merges onto a line of first-order phase transitions at a critical endpoint \cite{Stephanov:2004wx}. 

The transition temperature and the curvature of the pseudocritical line at small chemical potentials are known quite reliably from lattice simulations. The critical temperature turns out to be rather low ($T_{0}\simeq 157\,{\rm Mev}$ \cite{Bazavov:2018mes}), not much larger than the pion mass, and certainly smaller than any typical hadronic scale. At such low energies chiral perturbation theory should be a reasonable first approximation and, following \cite{Zarembo:2001wr}, we will estimate the shape of the whole critical line within this assumption, augmented with a simple model of chiral dynamics.
 
 The starting point is the standard chiral Lagrangian (neglecting quark masses to the first approximation):
\begin{equation}\label{chiL}
 \mathcal{L}_{\chi }=\frac{F^{2}}{4}\,\mathop{\mathrm{tr}}\partial _{\mu }U^\dagger \partial ^{\mu }U,\qquad U=\,{\rm e}\,^{iT^{a}\pi _{a}}.
\end{equation}
In this scenario the chiral condensate does not melt at the transition point but is rather phase-disordered. Indeed, the pion field can be regarded as the phase of the condensate: $\bar{\psi }\psi \sim \Sigma\,{\rm e}\,^{iT^{a}\pi _{a}} $,  and strong pion fluctuations, such that $\left\langle \mathop{\mathrm{tr}}\,{\rm e}\,^{iT^{a}\pi _{a}}\right\rangle=0$ will restore chiral symmetry even if the modulus of the condensate remains non-zero: $\left\langle\Sigma\right\rangle \neq 0$. This mechanism is physically different from the more standard scenario of gradually melting condensate, but the symmetry-breaking pattern is the same and universality considerations \cite{Pisarski:1983ms} apply here without change.

The chiral disorder can be triggered by proliferation of skyrmions \cite{Detar:1990tq,Kogan:2001ex}, in parallel to the Berezinskii-Kosterlitz-Thouless transition in 2d, or by a different physical mechanism 
\cite{Detar:1991cr,Bochkarev:1995gi,Babaev:1999in,Babaev:2000fj}. The details will not be important here, we just need to assume that $\left\langle \mathop{\mathrm{tr}}U\right\rangle\rightarrow 0$ at some critical temperature $T_{c}$, sufficiently low for the chiral Lagrangian to capture essential physics.
Dimensional analysis then dictates that
\begin{equation}\label{TF}
 T_{c}=\kappa F,
\end{equation}
where $\kappa $ is a numerical constant of order one\footnote{Taking $T_{0}=157\,{\rm Mev}$ and neglecting thermal corrections to the pion decay constant $F_{\pi }=92\,{\rm Mev}$ results in $\kappa =1.71$. A mean-field calculation in the $N_{f}=2$ non-linear sigma-model gives $\kappa =\sqrt{3}=1.73$ \cite{Bochkarev:1995gi}. So  perfect agreement is likely a numerical coincidence, and indeed thermal corrections (see below) diminish $F$ and hence raise $\kappa $ by $20\%$, but still the numbers remain within a reasonable range from one another.}.

To proceed further we need a model for the chiral Lagrangian itself. In principle, the chiral Lagrangian represents the low-energy approximation generated by integrating out heavy degrees of freedom in the full partition function of QCD while keeping the low-energy Goldstone modes as external fields.  For obvious practical reasons we consider a simpler phenomenological model amenable to analytical treatment \cite{Diakonov:1985eg,Aitchison:1986aq,Diakonov:1987ty,Diakonov:1997sj}:
\begin{equation}\label{q-1}
 \mathcal{L}=\bar{\psi }\left(\dc{\partial }+M\,{\rm e}\,^{i\gamma ^{5}T^{a}\pi _{a}}\right)\psi .
 \end{equation}
 The model describes the simplest coupling of Goldstone bosons to constituent quarks consistent with symmetries of QCD \cite{Manohar:1983md,Diakonov:1984tw,Dhar:1983fr,Dhar:1985gh}. The constituent quark mass can be estimated as a half of the $\rho $-meson mass or a third of the nucleon mass, and we will take $M=350^{+40}_{-40}\,{\rm Mev}$ in numerical estimates. 
 
 The effective action for Goldstone bosons is generated, within this model, by integrating out constituent quarks and expanding the resulting effective action in derivatives:
 \begin{equation}\label{Ch-q}
 S_{\chi }=-N_{c}\ln\det\left[\dc{\partial }+M\left(\frac{1+\gamma ^{5}}{2}\,U+\frac{1-\gamma ^{5}}{2}\,U^\dagger \right)\right].
\end{equation}
This model can be derived within the instanton liquid theory under the assumptions discussed in detail in \cite{Diakonov:1985eg,Diakonov:1995ea}. Simple as it is and having just one free parameter, the model yields reasonable values for low-energy constants of chiral perturbation theory \cite{Praszalowicz:1989dh} and underlies a quantitatively consistent description of nucleon as a chiral soliton \cite{Diakonov:1987ty,Diakonov:1997sj}. 

The chiral Lagrangian arises at the first order in the derivative expansion, with the pion decay constant given by \cite{Diakonov:1997sj}
\begin{equation}
 F^{2}=4N_{c}M^{2}\int_{}^{}\frac{d^{4}p}{\left(2\pi \right)^{4}}\,\,\frac{1}{\left(p^{2}+M^{2}\right)^{2}}\,.
\end{equation}
The $p_{0}$ integration at finite temperature and density should be understood as summation over Matsubara frequencies: $p^{0}= (2n-1)\pi T-i\mu $, where $\mu $ is the quark chemical potential. Standard manipulations with the Matsubara sum yield:
\begin{equation}\label{F2}
 F^{2}=F^{2}_{\pi }-\frac{N_{c}M^{2}}{2\pi ^{2}}\int_{M}^{\infty }\frac{d\omega }{\sqrt{\omega ^{2}-M^{2}}}
 \left(\frac{1}{\,{\rm e}\,^{\frac{\omega -\mu }{T}}+1}
 +\frac{1}{\,{\rm e}\,^{\frac{\omega +\mu }{T}}+1}
 \right),
\end{equation}
where  $F_{\pi }=92\,{\rm Mev}$. 

Pions do not carry baryon charge and the chemical potential enters the low-energy theory only through thermal corrections to the pion decay constant $F\equiv F(T,\mu )$.
Once the dependence of $F$ on $T$ and $\mu $ is known,
the critical line in the $T-\mu $ plane can be computed from (\ref{TF}):\begin{equation}\label{ccondition}
 \frac{F^{2}(T_{c}(\mu ),\mu )}{F^{2}(T_{0},0)}=\frac{T_{c}^{2}(\mu )}{T_{0}^{2}}\,.
\end{equation}
The shape of the critical line can be inferred from (\ref{ccondition}), (\ref{F2}). Since thermal fluctuations tend to diminish $F$, transition temperature will decrease with $\mu$, and the critical line will bend towards smaller temperatures at larger chemical potential: $T_{c}(\mu )< T_{0}$, as expected.

\begin{figure}[t]
\begin{center}
 \centerline{\includegraphics[width=12cm]{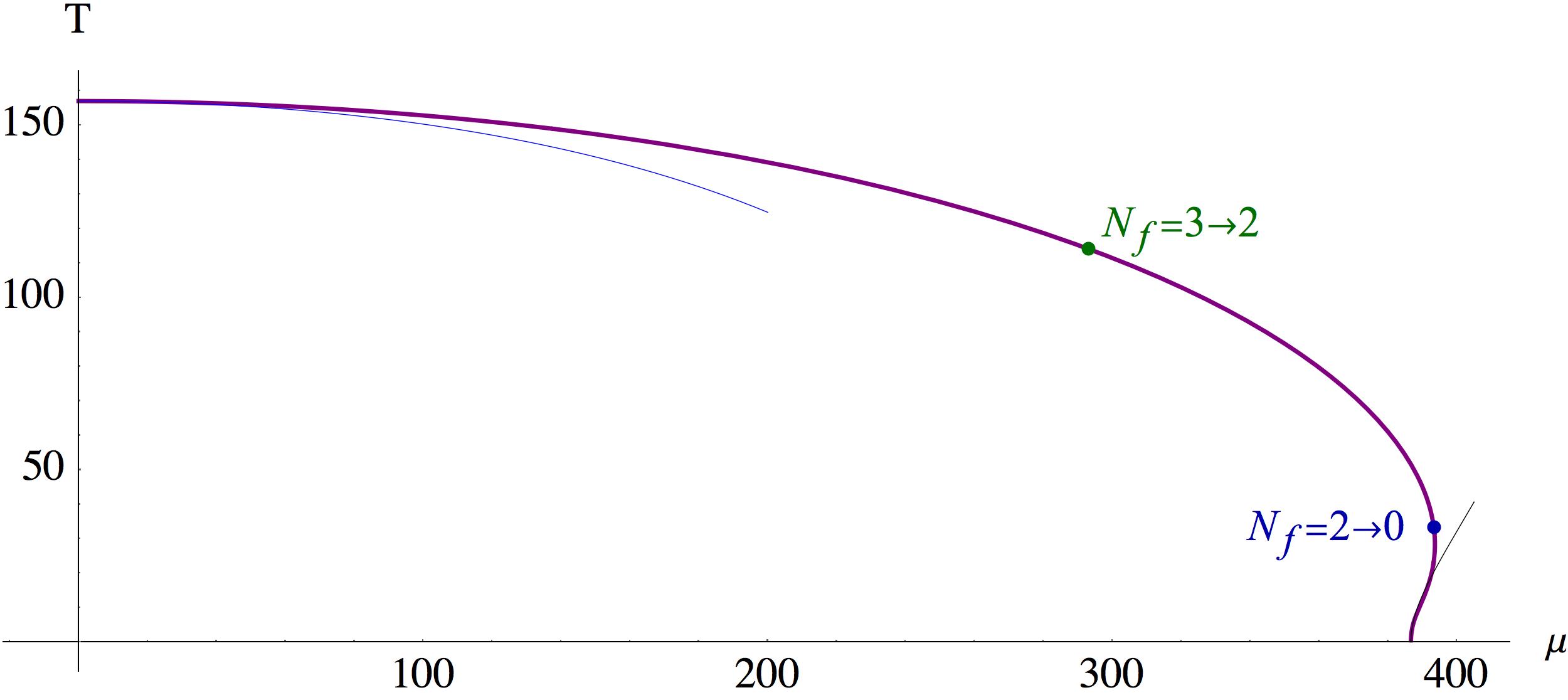}}
\caption{\label{PD}\small The pseudocritical line in the $T-\mu $ plane calculated from (\ref{ccondition}), (\ref{F2}). Boltzmann and Fermi approximations (\ref{BA})  and (\ref{Fermi}) are shown in thin lines. The dots indicate decoupling of kaons ($N_{f}=3\rightarrow 2$) and pions ($N_{f}=2\rightarrow 0$).}
\end{center}
\end{figure}

The predicted critical line is shown in fig.~\ref{PD}. It starts very flat, abruptly curves down at $\mu \sim M$ and passes through a turning point before terminating  at $(T,\mu )=(0,\mu _{*})$ with $ \mu _{*}=387\,{\rm Mev}$. This behavior is easily understood at the qualitative level. 

Consider first small chemical potentials. The relevant parameter then is  the ratio $M/T_{0}\simeq 2.2$, which is not huge but an exponential dependence on $M/T$ leads to considerable, $\mathcal{O}(10)$ Boltzmann suppression. Expanding the integral  (\ref{F2})  in  $T/M$, one gets:
\begin{equation}\label{FBoltzmann}
 F^{2}\simeq F_{\pi }^{2}-\frac{N_{c}M^{2}}{\pi }\sqrt{\frac{T}{2\pi M}}\,
 \,{\rm e}\,^{-\frac{M}{T}}\cosh\frac{\mu }{T}\,,
\end{equation}
which yields a simple approximate formula for the critical line \cite{Zarembo:2001wr}:
\begin{equation}\label{BA}
 \frac{T_{c}^{2}}{T_{0}^{2}}\simeq 1-\frac{N_{c}M^{2}}{\pi F_{\pi }^{2}}\sqrt{\frac{T_{0}}{2\pi M}}\,
 \,{\rm e}\,^{-\frac{M}{T_{0}}}\left(\cosh\frac{\mu }{T_{0}}-1\right).
\end{equation}
As seen from fig.~\ref{PD}, Boltzmann approximation  works reasonably well as long as $\mu \lesssim 100\,{\rm Mev}$. 

The critical temperature at small $\mu $ is often represented by Taylor expansion in the baryon chemical potential:
\begin{equation}
 \frac{T_{c}}{T_{0}}=1-\kappa _{2}\,\frac{\mu_{B} ^{2}}{T_{0}^{2}}-\kappa _{4}\,\frac{\mu _{B}^{4}}{T_{0}^{4}}+\ldots ,
\end{equation}
where $\mu _{B}= N_{c}\mu$. From (\ref{BA}) we get for the first two coefficients:
\begin{eqnarray}
 \kappa _{2}&=&\frac{M^{2}}{4\pi N_{c}F_{\pi }^{2}}\sqrt{\frac{T_{0}}{2\pi M}}\,
 \,{\rm e}\,^{-\frac{M}{T_{0}}}=0.0111^{+0.0008}_{-0.0010}
\nonumber \\
\kappa _{4}&=&\frac{\kappa _{2}^{2}}{2}+\frac{M^{2}}{24\pi N^{3}_{c}F_{\pi }^{2}}\sqrt{\frac{T_{0}}{2\pi M}}\,
 \,{\rm e}\,^{-\frac{M}{T_{0}}}=0.00027^{+0.00002}_{-0.00003}.
\end{eqnarray}
The error intervals are obtained by varying $M$ between $310$ and $390\,{\rm Mev}$. These estimates are in reasonable agreement with recent lattice results: $\kappa _{2}=0.012(4)$ and $\kappa _{4}=0.000(4)$ \cite{Bazavov:2018mes}\footnote{There are several lattice \cite{DElia:2018fjp} and phenomenological \cite{Becattini:2016xct} determinations of the curvature of the critical line which are broadly consistent with each other \cite{Cea:2014xva,Bonati:2016fbi}.}. The tiny value of $\kappa _{2}$ and the ensuing flatness of the critical line here are simple consequences of the Boltzmann suppression.

In the opposite limit, when temperature goes to zero, the Fermi distribution becomes a step function, and the integral in (\ref{F2}) can again be computed analytically. With the first temperature correction taken into account,
\begin{equation}\label{Fermi}
 F^{2}\simeq F_{\pi }^{2}-\frac{N_{c}M^{2}}{4\pi ^{2}}\,
 \ln\frac{\mu +\sqrt{\mu ^{2}-M^{2}}}{\mu -\sqrt{\mu ^{2}-M^{2}}}
 +\frac{N_{c}\mu M^{2}T^{2}}{12(\mu ^{2}-M^{2})^{\frac{3}{2}}}\,.
\end{equation}

The decay constant at zero temperature, $F(0,\mu )$, diminishes with $\mu $ and eventually turns to zero at $\mu =\mu _{*}$, where 
\begin{equation}
 \mu _{*}=M\cosh\frac{2\pi ^{2}F_{\pi }^{2}}{N_{c}M^{2}}=387\,{\rm Mev}.
\end{equation}
This is exactly where the critical line terminates, because (\ref{ccondition}) implies that $T_{c}(\mu _{*})=0$ once $F(0,\mu _{*})=0$. To continue the critical line into the domain of small but non-zero temperatures,  one has to keep the $\mathcal{O}(T^{2})$ term in (\ref{Fermi}), which yields for the critical temperature:
\begin{equation}
 \frac{T_{c}^{2}}{T_{0}^{2}}\simeq \frac{\frac{N_{c}M^{2}}{4\pi ^{2}F_{\pi }^{2}}\,
 \ln\frac{\mu +\sqrt{\mu ^{2}-M^{2}}}{\mu -\sqrt{\mu ^{2}-M^{2}}}-1}{\frac{N_{c}\mu M^{2}T_{0}^{2}}{12F_{\pi }^{2}(\mu ^{2}-M^{2})^{\frac{3}{2}}}-1}
\end{equation}
 As seen from fig.~\ref{PD}  this is a good approximation only for very low temperatures $T\lesssim 20\,{\rm Mev}$. The critical curve then almost exactly follows the $F^{2}(T,\mu )=0$ line, due to the numerical proximity of $\mu _{*}$ to $M$ that enhances the $\mathcal{O}(T^{2})$ coefficient in (\ref{Fermi}). The small denominator and the positive sign of the $\mathcal{O}(T^{2})$ correction explain why the critical curve makes a wiggle and has to pass through a turning point to match the declining bulk segment.

The current quark masses have been completely neglected so far. Once included,  
\begin{equation}
 S_{\chi }=-N_{c}\ln\det\left[\dc{\partial }+m_{q}+M\left(\frac{1+\gamma ^{5}}{2}\,U+\frac{1-\gamma ^{5}}{2}\,U^\dagger \right)\right],
\end{equation}
they generate Goldstone boson masses at the linear order in $m_{q}$:
\begin{equation}
 \mathcal{L}_{\chi }=\frac{F^{2}}{4}\,\mathop{\mathrm{tr}}\left[\partial _{\mu }U^\dagger \partial ^{\mu }U-m_{{\rm eff}}^{2}\left(U+U^\dagger \right)\right],
\end{equation}
where
\begin{equation}
 F^{2}m^{2}_{{\rm eff}}=8N_{c}Mm_{q}\int_{}^{}\frac{d^{4}p}{\left(2\pi \right)^{4}}\,\,\frac{1}{p^{2}+M^{2}}\,.
\end{equation}
The last formula is equivalent to the Gell-Mann-Oaks-Reiner relation, and at finite temperature and density reads
\begin{equation}\label{effmass}
 F^{2}m_{{\rm eff}}^{2}=F_{\pi }^{2}m_{G}^{2}-\frac{2N_{c}Mm_{q}}{\pi ^{2}}
 \int_{M}^{\infty }d\omega \,\sqrt{\omega ^{2}-m^{2}}
  \left(\frac{1}{\,{\rm e}\,^{\frac{\omega -\mu }{T}}+1}
 +\frac{1}{\,{\rm e}\,^{\frac{\omega +\mu }{T}}+1}
 \right).
\end{equation}
Here $m_{G}^{2}$ is the pseudo-Goldstone mass-squared matrix at zero temperature and density. The mass eigenvalues are computed by substituting $m_{q}=m_{u}+m_{d}$ for $\pi $, $m_{q}=m_{s}+m_{u}$ for $K^{\pm}$, $m_{q}=m_{s}+m_{d}$ for $K^{0}$, and $m_{q}=(4m_{s}+m_{d}+m_{u})/3$ for $\eta $. 

The dependence on the chemical potential is again oblique, allowing for extrapolation of zero-density data into the whole $\mu -{T}$ plane, as was done before in computing the critical line. For physical values of the quark masses and at zero chemical potential the transition is known to become a crossover, but once $m_{{u,d,s}}$ and hence $m _{\pi }^{2}$ and $m_{K}^{2}$ are allowed to vary, the nature of the transition may change. A first-order transition is expected in the very corner of the $m_{\pi }^{2}-m_{K}^{2}$ diagram, known as the Columbia plot \cite{Brown:1990ev} (see \cite{deForcrand:2017cgb,Cuteri:2018wci} for a summary of recent lattice results).  Turning on the chemical potential leads to variation in the Goldstone boson masses according to (\ref{effmass}). Thermal corrections effectively move the masses of kaons and pions away from the physical point on the Columbia plot and one can imagine that eventually they could reach the phase-transition domain. Crossing the separatrix between crossover and first-order phase transition on the Columbia plot would then correspond to the critical endpoint on the $T-\mu $ plane. 

But this assumes that Goldstone masses decrease along the pseudocritical line. This assumptions turns out to be wrong. At very small chemical potentials one can use Boltzmann approximation in (\ref{effmass}). Expanding in $T/M$ and using (\ref{FBoltzmann}) for the pion decay constant, we get:
\begin{equation}
 m_{{\rm eff}}^{2}\simeq m_{G}^{2}
 \left[1+\frac{N_{c}M^{2}}{\pi F_{\pi }^{2}}
 \sqrt{\frac{T}{2\pi M}}\,\,{\rm e}\,^{-\frac{M}{T}}\cosh\frac{\mu }{T}\,
 \left(1-\frac{6Tm_{q}}{m_{G}^{2}}\right)\right].
\end{equation}
 Since $6T_{0}m_{q}/m_{G}^{2}=0.36$, the correction drives the masses upwards. The mass corrections become even larger at lower temperatures because $F(T_{c},\mu )$ turns to zero at the endpoint of the critical curve, while the right-hand side of (\ref{effmass}) remains finite.
 
 \begin{figure}[t]
\begin{center}
 \centerline{\includegraphics[width=8cm]{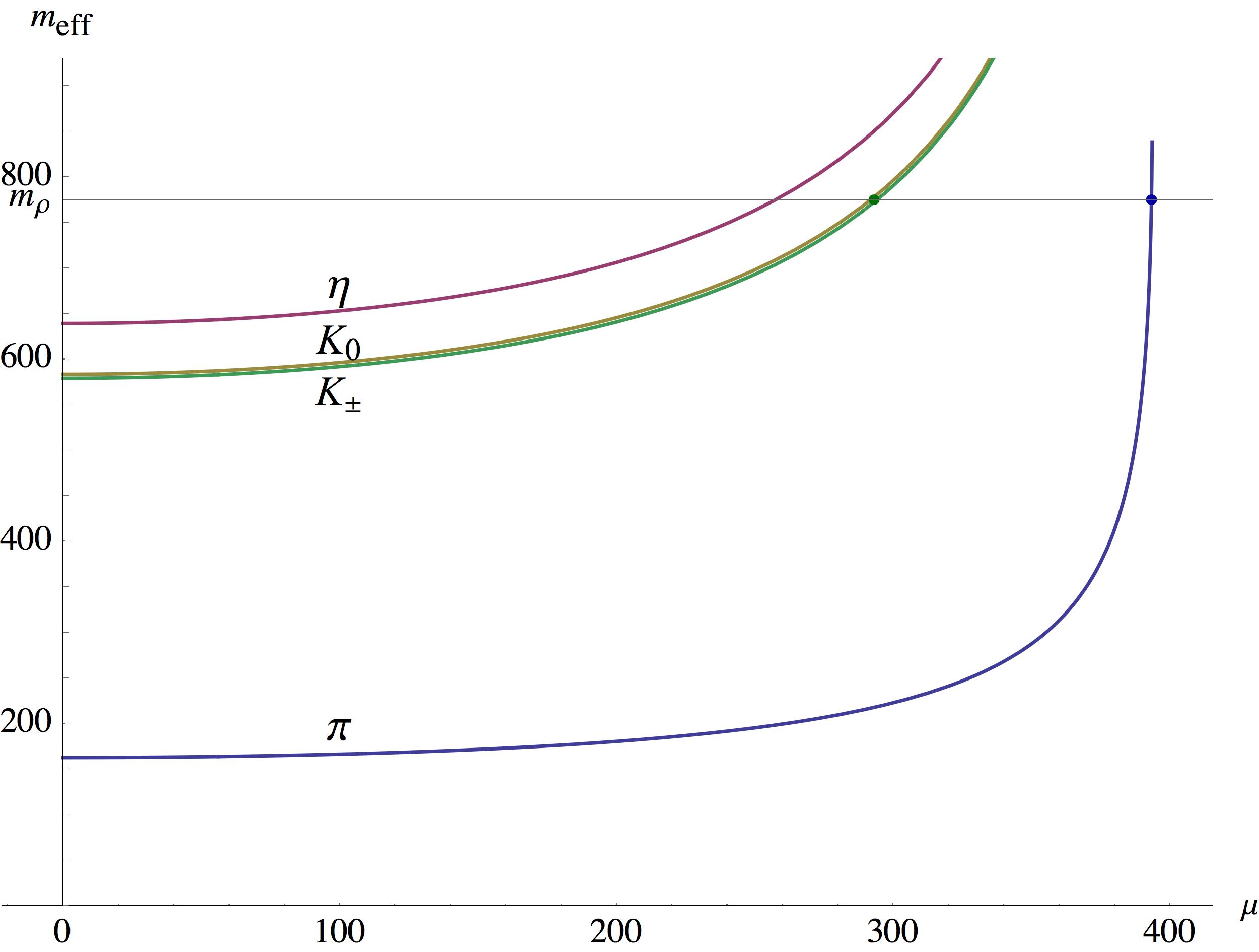}}
\caption{\label{GM}\small Effective masses of the pseudo-Goldstone bosons along the pseudocritical line, $m_{{\rm eff}}(T_{c}(\mu ),\mu )$. }
\end{center}
\end{figure}

 The variation of the Goldstone boson masses along the pseudocritical line,  as computed from (\ref{effmass}), is shown in fig.~\ref{GM}. The masses grow all along and eventually kaons and pions become indistinguishable from heavier, non-Goldtone mesons. Somewhat arbitrarily, we take equality to the $\rho $-meson mass as the decoupling condition.  The points where kaons and pions decouple are marked in fig.~\ref{GM} and also on the phase diagram in fig.~\ref{PD}. Since the Goldstone boson masses increase along the crossover line the critical endpoint never arises in this scenario. The crossover in fact becomes weaker and weaker at larger chemical potentials. This is perhaps the most interesting qualitative prediction of this model.  

We considered a scenario in which the chiral crossover is triggered by strong pion fluctuations which disorder the chiral condensate's phase.  This mechanism is physically distinct from gradual melting of the condensate and should leave characteristic imprints on the observables close to the transition point \cite{Detar:1990tq,Detar:1991cr}. One distinct feature of this scenario is an absence of the critical endpoint on the phase diagram. Crossover always remains a crossover, and moreover becomes less sharp with growing chemical potential. This relatively unconventional conclusion is consistent with recent lattice simulations of the $N_{c}=2$, $N_{f}=2$ theory \cite{Bornyakov:2017txe}, where no sharp transition was observed.

Simple as it is the model gives concrete predictions for the shape of the pseudo-critical line in the $\mathbf{\mu }-T$ plane, whose curvature at small chemical potentials is in reasonable agreement with recent lattice evaluations. It would be interesting to compute other quantities within this model, for example temperature dependence of the chiral condensate or the cumulants of the baryon number fluctuations.
 
\subsection*{Acknowledgements}
We would like to thank V.~Braguta, D.~Kharzeev, E.~Langmann and A.~Mo\-loch\-kov for interesting discussions and to A.~Kotov for email correspondence. This work of was supported by the Swedish Research Council (VR) grant
2013-4329, by the grant "Exact Results in Gauge and String Theories" from the Knut and Alice Wallenberg foundation, and by RFBR grant 18-01-00460 A. 

\appendix

\bibliographystyle{nb}

\begin{thebibliography}{10}
\ifx\href\asklfhas\newcommand{\href}[2]{#2}\fi
\raggedright
\small
\parskip 0pt

\bibitem{Stephanov:2004wx}
M.~A.~Stephanov,
\textit{``{QCD phase diagram and the critical point}''},
\textsf{Prog.~Theor.~Phys.~Suppl.~153,~139~(2004)},
\href{http://arXiv.org/abs/hep-ph/0402115}{\texttt{hep-ph/0402115}}.
%
\bibitem{Bazavov:2018mes}
A.~Bazavov et~al.,
\textit{``{Chiral crossover in QCD at zero and non-zero chemical
  potentials}''},
\href{http://arXiv.org/abs/1812.08235}{\texttt{1812.08235}}.
%
\bibitem{Zarembo:2001wr}
K.~Zarembo,
\textit{``{Possible pseudogap phase in QCD}''},
\textsf{JETP~Lett.~75,~59~(2002)},
\href{http://arXiv.org/abs/hep-ph/0104305}{\texttt{hep-ph/0104305}}.
%
\bibitem{Pisarski:1983ms}
R.~D.~Pisarski and F.~Wilczek,
\textit{``{Remarks on the Chiral Phase Transition in Chromodynamics}''},
\textsf{Phys.~Rev.~D29,~338~(1984)}.
%
\bibitem{Detar:1990tq}
C.~E.~Detar,
\textit{``{The Role of Baryons in Chiral Symmetry Restoration at High
  Temperature}''},
\textsf{Phys.~Rev.~D42,~224~(1990)}.
%
\bibitem{Kogan:2001ex}
I.~I.~Kogan, A.~Kovner and B.~Tekin,
\textit{``{The chicken or the egg: Or Who ordered the chiral phase
  transition?}''},
\textsf{Phys.~Rev.~D63,~116007~(2001)},
\href{http://arXiv.org/abs/hep-ph/0101040}{\texttt{hep-ph/0101040}}.
%
\bibitem{Detar:1991cr}
C.~E.~Detar and S.-J.~Dong,
\textit{``{Toward an effective chiral model of high temperature QCD}''},
\textsf{Phys.~Rev.~D45,~4681~(1992)}.
%
\bibitem{Bochkarev:1995gi}
A.~Bochkarev and J.~I.~Kapusta,
\textit{``{Chiral symmetry at finite temperature: Linear versus nonlinear sigma
  models}''},
\textsf{Phys.~Rev.~D54,~4066~(1996)},
\href{http://arXiv.org/abs/hep-ph/9602405}{\texttt{hep-ph/9602405}}.
%
\bibitem{Babaev:1999in}
E.~Babaev,
\textit{``{Mass generation without symmetry breakdown in the chiral Gross-Neveu
  model at finite temperature and finite N in (2+1)-dimensions}''},
\textsf{Phys.~Lett.~B497,~323~(2001)},
\href{http://arXiv.org/abs/hep-th/9907089}{\texttt{hep-th/9907089}}.
%
\bibitem{Babaev:2000fj}
E.~Babaev,
\textit{``{Nonlinear sigma model approach for chiral fluctuations and symmetry
  breakdown in Nambu-Jona-Lasinio model}''},
\textsf{Phys.~Rev.~D62,~074020~(2000)},
\href{http://arXiv.org/abs/hep-ph/0006087}{\texttt{hep-ph/0006087}}.
%
\bibitem{Diakonov:1985eg}
D.~Diakonov and V.~{\relax Yu}.~Petrov,
\textit{``{A Theory of Light Quarks in the Instanton Vacuum}''},
\textsf{Nucl.~Phys.~B272,~457~(1986)}.
%
\bibitem{Aitchison:1986aq}
I.~Aitchison, C.~Fraser, E.~Tudor and J.~Zuk,
\textit{``{Failure of the Derivative Expansion for Studying Stability of the
  Baryon as a Chiral Soliton}''},
\textsf{Phys.~Lett.~165B,~162~(1985)}.
%
\bibitem{Diakonov:1987ty}
D.~Diakonov, V.~{\relax Yu}.~Petrov and P.~V.~Pobylitsa,
\textit{``{A Chiral Theory of Nucleons}''},
\textsf{Nucl.~Phys.~B306,~809~(1988)}.
%
\bibitem{Diakonov:1997sj}
D.~Diakonov,
\textit{``{Chiral quark - soliton model}''},
\href{http://arXiv.org/abs/hep-ph/9802298}{\texttt{hep-ph/9802298}}.
%
\bibitem{Manohar:1983md}
A.~Manohar and H.~Georgi,
\textit{``{Chiral Quarks and the Nonrelativistic Quark Model}''},
\textsf{Nucl.~Phys.~B234,~189~(1984)}.
%
\bibitem{Diakonov:1984tw}
D.~Diakonov and M.~I.~Eides,
\textit{``{Chiral lagrangian from a functional integral over quarks}''},
\textsf{JETP~Lett.~38,~433~(1983)}.
%
\bibitem{Dhar:1983fr}
A.~Dhar and S.~R.~Wadia,
\textit{``{The Nambu-Jona-Lasinio Model: An Effective Lagrangian for Quantum
  Chromodynamics at Intermediate Length Scales}''},
\textsf{Phys.~Rev.~Lett.~52,~959~(1984)}.
%
\bibitem{Dhar:1985gh}
A.~Dhar, R.~Shankar and S.~R.~Wadia,
\textit{``{Nambu-Jona-Lasinio Type Effective Lagrangian. 2. Anomalies and
  Nonlinear Lagrangian of Low-Energy, Large N QCD}''},
\textsf{Phys.~Rev.~D31,~3256~(1985)}.
%
\bibitem{Diakonov:1995ea}
D.~Diakonov,
\textit{``{Chiral symmetry breaking by instantons}''},
\textsf{Proc.~Int.~Sch.~Phys.~Fermi~130,~397~(1996)},
\href{http://arXiv.org/abs/hep-ph/9602375}{\texttt{hep-ph/9602375}}.
%
\bibitem{Praszalowicz:1989dh}
M.~Praszalowicz and G.~Valencia,
\textit{``{Quark Models and Chiral Lagrangians}''},
\textsf{Nucl.~Phys.~B341,~27~(1990)}.
%
\bibitem{DElia:2018fjp}
M.~D'Elia,
\textit{``{High-Temperature QCD: theory overview}''},
\textsf{Nucl.~Phys.~A982,~99~(2019)},
\href{http://arXiv.org/abs/1809.10660}{\texttt{1809.10660}}.
%
\bibitem{Becattini:2016xct}
F.~Becattini, J.~Steinheimer, R.~Stock and M.~Bleicher,
\textit{``{Hadronization conditions in relativistic nuclear collisions and the
  QCD pseudo-critical line}''},
\textsf{Phys.~Lett.~B764,~241~(2017)},
\href{http://arXiv.org/abs/1605.09694}{\texttt{1605.09694}}.
%
\bibitem{Cea:2014xva}
P.~Cea, L.~Cosmai and A.~Papa,
\textit{``{Critical line of 2+1 flavor QCD}''},
\textsf{Phys.~Rev.~D89,~074512~(2014)},
\href{http://arXiv.org/abs/1403.0821}{\texttt{1403.0821}}.
%
\bibitem{Bonati:2016fbi}
C.~Bonati, M.~D'Elia, M.~Mariti, M.~Mesiti, F.~Negro and F.~Sanfilippo,
\textit{``{Across the deconfinement}''},
\textsf{Acta~Phys.~Polon.~Supp.~10,~489~(2017)},
\href{http://arXiv.org/abs/1610.03338}{\texttt{1610.03338}}.
%
\bibitem{Brown:1990ev}
F.~R.~Brown, F.~P.~Butler, H.~Chen, N.~H.~Christ, Z.-h.~Dong, W.~Schaffer,
  L.~I.~Unger and A.~Vaccarino,
\textit{``{On the existence of a phase transition for QCD with three light
  quarks}''},
\textsf{Phys.~Rev.~Lett.~65,~2491~(1990)}.
%
\bibitem{deForcrand:2017cgb}
P.~de~Forcrand and M.~D'Elia,
\textit{``{Continuum limit and universality of the Columbia plot}''},
\textsf{PoS~LATTICE2016,~081~(2017)},
\href{http://arXiv.org/abs/1702.00330}{\texttt{1702.00330}}.
%
\bibitem{Cuteri:2018wci}
F.~Cuteri, O.~Philipsen and A.~Sciarra,
\textit{``{Progress on the nature of the QCD thermal transition as a function
  of quark flavors and masses}''},
\href{http://arXiv.org/abs/1811.03840}{\texttt{1811.03840}}.
%
\bibitem{Bornyakov:2017txe}
V.~G.~Bornyakov, V.~V.~Braguta, E.~M.~Ilgenfritz, A.~{\relax Yu}.~Kotov,
  A.~V.~Molochkov and A.~A.~Nikolaev,
\textit{``{Observation of deconfinement in a cold dense quark medium}''},
\textsf{JHEP~1803,~161~(2018)},
\href{http://arXiv.org/abs/1711.01869}{\texttt{1711.01869}}.
%
\end{thebibliography}

\end{document}